\documentclass[12pt,aps,floats,superscriptaddress,showkeys]{revtex4-1}
\usepackage{latexsym}
\usepackage{amsmath}
\usepackage{amsfonts}
\usepackage{amssymb}
\usepackage{graphicx}
\usepackage{braket}
\usepackage{comment}
\usepackage{graphicx}
 \usepackage{float}
\usepackage{subfig}
\usepackage{titlesec}
\usepackage{setspace}
\usepackage{breqn}
\usepackage{color}
\newcommand\beq{\begin{equation}}      \newcommand\eeq{\end{equation}}
\newcommand\bea{\begin{eqnarray}}      \newcommand\eea{\end{eqnarray}}
      \date{\today}
\usepackage[margin=1in]{geometry} 

\makeatletter
\let\cat@comma@active\@empty
\makeatother
\begin{document}

\title{\bf A Study of Asymptotic Freedom like Behavior for 
Topological States of Matter}

\author{Ranjith  Kumar   R}  \affiliation{Poornaprajna   Institute  of
  Scientific  Research, 4,  16th Cross,  Sadashivnagar, \\Bengaluru  -
  5600-80, India.}  \affiliation{Manipal  Academy of Higher Education,
  Madhava  Nagar,  \\Manipal  -   576104,  India.}   \author{Rahul  S}
\affiliation{Poornaprajna  Institute of  Scientific Research,  4, 16th
  Cross,    Sadashivnagar,    \\Bengaluru     -    5600-80,    India.}
\affiliation{Manipal  Academy  of  Higher  Education,  Madhava  Nagar,
  \\Manipal    -     576104,    India.}      \author{Surya    Narayan}
\affiliation{Raman Research Institute, C.  V. Raman Avenue, 5th Cross,
  Sadashivanagar, \\Bengaluru - 5600-80, India.}  \author{Sujit Sarkar
}  \email{E-mail   -  sujit.tifr@gmail.com}  \affiliation{Poornaprajna
  Institute  of Scientific  Research,  4,  16th Cross,  Sadashivnagar,
  \\Bengaluru - 5600-80, India.}

 \begin{abstract}
 \noindent   
 We
 present  results  for  asymptotic   freedom  like  behavior  for  the
 topological  state  of  the  helical spin liquid system with finite  proximity  induced
 superconducting gap.  
 We derive two different quantum Berezinskii-Kosterlitz-Thouless  (BKT) 
 equations for the two different limit of this model Hamiltonian. 
The common quantum phase for these two quantum BKT transitions is the 
helical Luttinger liquid phase where there is no evidence of asymptotic
freedom. There is no evidence of superconductor-insulator transition for 
this asymptotic freedom study. We observe the evidance of asymptotic freedom for the two model Hamiltonian, but the character of the asymptotic phases are different. We also observe that the Luttinger liquid parameter plays a significant role to determine the asymptotic freedom but the chemical potential has no effect on it.
 \end{abstract}

   \maketitle
	
	\textbf{Introduction}\\
	\noindent In the  year 1955, Landau identified  the problem that
        in the  presence of an  arbitrary number  of virtual  
        particles and a
        real  particle  there  will  be  no  interaction  due  to 
        screening \cite{landau1955point}.  This  problem was solved by
        David Gross and Frank Wilczek
        \cite{wilczek2005asymptotic,PhysRevLett.30.1343}  and
        independently   by   David   Politzer   in   the   same   year
        \cite{PhysRevLett.30.1346},  by finding  anti-screening effect
        or asymptotic behavior  in nonabelian gauge  theories, or
        Yang-Mills theories  \cite{Yang-Mill}. This  is the  origin of
        Quantum  Chromodynamics  (QCD),  which is  the  quantum  field
        theory of  strong interactions. This theory  explains that the
        coupling constant  between two quarks increases  as the length
        scale increases, and the  coupling decreases asymptotically as
        the  length scale  decreases,  making  it asymptotically  free
        theory  \cite{GellMann:1964nj,Zweig:1981pd}.   The search  for
        individual quarks  was not successful  and they were always  found in
        the bound state of quark and anti-quark or the bound states
        of three quarks
        \cite{nambu1965proceedings,han1965three,greenberg1964spin}.
        This principle of confinement was contradicted by the  observation of an
        interesting  fact by  the SLAC  (Stanford Linear  Accelerator)
        experiment   of  colliding   proton  with   energetic  photons
        \cite{friedman1972deep}.\\  
        In condensed  matter field  theory
        one often encounters the theories with
        divergences. Renormalization group technique  was used to deal
        with  these   divergences  and  to  get   finite  results  for
        observable  physical   quantities in  the  theory
        \cite{altland2010condensed,fradkin2013field,marino2017quantum,
         giamarchi2003quantum}. Using
        renormalization theory Feynman,  Schwinger, and Tomonoga wrote
        down  the  corrections due  to  interactions  with any  finite
        number       of      virtual       particles      in       QED
        \cite{feynman1948relativistic,schwinger1948quantum,tomonaga1946}. 
        It
        was extended to a wider class of theories  by G. 't  Hooft and
        M. Veltman \cite{hooft1972}.   To simplify the renormalization
        group  analysis,  in 1970,  Curtis  Callan  and Kurt  Symanzik
        derived  a  differential equation  (Callan-Symanzik  equation)
        which determines  change in the n-point  correlation functions
        under     variation      of     the      scaling     parameter
        \cite{callan1970broken,symanzik1970small}. The running
        coupling  constant  or  the   correlation  function  which  we
        encounter in the theories depends  on the momentum scale. This
        happens in a well-defined way  through $\beta$ function in the
        Callan-Symanzik   equation   \cite{shankar2017quantum}.    The
        $\beta$ function also gives the  behavior of coupling at large
        momentum to describe  asymptotic freedom in  QCD and at small
        momentum to describe critical phenomena. Physical systems
        which do not  have natural intrinsic energy  scale will remain
        neutral for any variation in the energy scales. This intrinsic
        energy  scale for  any  system is  the  one which  determines
        whether the  system is in  strong or weak  interaction. Apart
        from  these  there  are systems  which  possess  characteristic
        energy scale which  acts as the separation  between strong and
        weak  interaction  regimes which  helps  one  to identify  the
        asymptotic nature of the system. These asymptotic behaviors can
        be  witnessed  in condensed  matter  systems  such  as,
        Gross-Neveu model  of polyacetylene \cite{gross1974dynamical},
        2d   $\sigma$-model   \cite{non-linear-sigma},   Kondo   model
        \cite{kondo1964resistance}, superinsulating state of superconductor-insulator 
        transition in superconducting films \cite{SI-nature} etc.   In all these models  one can
        calculate the  Callan-Symanzik equation  and also  the $\beta$
        function  to know  the dependence  of correlation  function or
        coupling constant on the different energy scales. The negative
        $\beta$ function which  has been obtained in  all these models
        implies that  at low energy  scales the effective  coupling is
        large  and  system is  strongly  interacting,  and at high  energy
        scales  the  effective coupling  is  small  making the  system
        asymptotically free  \cite{marino2017quantum}.  The asymptotic
        freedom for  topological insulators has not  been studied yet,
        and it is still an open problem. We will show explicitly that the concept of asymptotic freedom like behavior observed in high energy physics, can also emerge in the quantum many body condensed matter system with topological background. \\ Another important application
        of   renormalization    group   (RG)   method   is    in   the
        Berezinskii-Kosterlitz-Thouless        (BKT)        transition
        \cite{jose2017duality,2013ybkt.book...93O,bietenholz2016berezinskii}. 
         Berezinskii
        \cite{berezinskii1971destruction} in  1971 and  Kosterlitz and
        Thouless \cite{KT-1973} in 1973  explained a topological phase
        transition in two dimensional XY spin model, since there is no
        possibility of spontaneous symmetry breaking  for $d \le 2$ (d
        is the dimension)  \cite{mermin1966absence,PhysRev.158.383}.   The
        study  of  BKT transition  is  crucial  in quantum  many  body
        systems since many quantum  mechanical two dimensional systems
        can   be   approximated   to    two   dimensional   XY   model
        \cite{benfatto2013berezinskii}. In this study, we would like to unify two
                different regimes of RG theory, the asymptotic 
                freedom like  behavior of high-energy physics and the
                BKT transition of condensed matter physics.\\   
        Here    we   consider   an
        interacting helical  liquid system  at the  edge as  our model
        Hamiltonian. The quantum spin Hall systems are associated with
        the  helical  liquid  system which  describes  the  connection
        between  spin and  momentum. The  left movers  in the  edge of
        quantum spin  Hall systems are  associated with down  spin and
        right movers with up spin
        \cite{PhysRevLett.95.146802,topo.insu,moore2010birth,nishimori2010elements,
PhysRevB.80.155131,PhysRevB.85.075125,PhysRevB.83.035107}. The
        helical liquid  system possess  the gapless excitation  at the
        edge and  this results  in the
        appearance of  Majorana particles at  both ends of  the system
        \cite{Sarkar-2016}. \\We present 
        the motivation for this study below.\\ 
\textbf{Motivation and importance of this study:} \\
{\bf First objective}: 
The physics of topological state of matter is the second
revolution of quantum mechanics \cite{haldane}. This important concept and new important
results with high impact not only bound to the general audience of different
branches of physics but also creates interest for the other branches of science (Mathematics, Chemistry, Biology, Engineering and also in the Philosophy of Science). 
This new revolution in quantum mechanics was honored by the Nobel 
prize in physics in the year 2016.
This is 
one of the fundamental motivation of this present study. \\
{\bf Second objective}:
The mathematical structure and results of the renormalization group (RG) theory are the
most significant conceptual advancement in quantum field theory in the last several decades
in both high-energy and condensed matter physics.
The need for RG is really 
transparent in condensed matter physics. RG theory is a formalism that relates the physics at
different length scales in condensed matter physics and the physics at different energy scales
in high energy physics.
In the present study, we show explicitly how the physics of topology appears at different
length scales of the system and also obtain the signature of asymptotic freedom like 
behaviour. \\
{\bf Third objective}: The most important success of the RG
theory is the prediction of Asymptotic freedom (Physics Nobel Prize 2004) and the physics
of BKT (Physics Nobel Prize 2016), although it is classical BKT (the present problem
is on quantum BKT which is more subtle than classical BKT). The 
concept of asymptotic freedom is not only confined to high energy physics but also we show that  it is present in quantum condensed matter systems. In the present work, we
unified these two concepts in a single framework of topological state of matter. There are
no such studies in the entire literature of topological state of matter. These new,
important and original results will provide a new perspective on the study of topological
state of matter. \\
{\bf Fourth objective}:
         Here we mention very briefly the nature of different Luttinger 
         liquid physics to emphasise the rich physics of helical Luttinger 
         liquid. 
        The physics of Luttinger liquids (LL) can be of three different 
        forms : spinful LL, chiral LL, and helical LL. Spinful LL shows 
        linear dispersion around the fermi level with the difference of 
        $2k_F$, in the momentum between left and right moving branches. 
        Chiral LL has spin degenerated, strongly correlated electrons 
        moving in only one direction. In helical LL one can observe the 
        Dirac point due to the crossing of left and right moving branches, 
        also electrons with opposite spins move in opposite directions.
        We consider this physics for the present model Hamiltonian system. 
        The Luttinger parameter ($K$) determines
         the nature of interaction. $K<1$ and $K>1$ characterize the 
         repulsive and attractive interactions respectively, whereas $K=1$
         characterize the non-interacting situation. The present study show
         the importance of $K$ for quantum phases and asymptotic freedom  
        \cite{helical1}.\\\\
        \textbf{Model Hamiltonian}\\
	\noindent We consider the interacting helical liquid system at
        the  edge  of a  topological  insulator  as our  model  system
        \cite{PhysRevLett.95.146802}. These edge  states are protected
        by  the  symmetries  \cite{PhysRevB.85.075125}. In the edge
        states of helical  liquid, spin and momentum  are connected as
        the  right movers  are associated  with the  spin up  and left
        movers are  associated with spin down. 
        One can  write the
                total fermionic  field of the system  as $$\psi(x)=e^{ik_F x}
                \psi_{R\uparrow}+e^{-ik_F   x}   \psi_{L\downarrow},$$   where
                $\psi_{R\uparrow}$  and  $\psi_{L\downarrow}$  are  the  field
                operators  corresponding to  right moving  (spin up)  and left
                moving (spin down) electron at  the upper and lower edges
                of the topological insulators \cite{Altland-2011,wu2006helical,Sarkar-2016,PhysRevLett.107.036801}.
               For the  low energy  collective excitation in  one dimensional
                system one can write the Hamiltonian as
        	\begin{equation} 
        		H_0  =  \int  \frac{dk}{2  \pi}  {v_F}  [  ({{\psi}_{R
                              \uparrow}}^{\dagger}  (i   \partial_x)  {\psi_{R
                              \uparrow} }  - {{\psi}_{L \downarrow}}^{\dagger}
                          (i   \partial_x)  {\psi_{L   \downarrow}  })   \\  +
                          ({{\psi}_{R  \downarrow}}^{\dagger}  (i  \partial_x)
                          {\psi_{R     \downarrow}      }     -     {{\psi}_{L
                              \uparrow}}^{\dagger}  (i   \partial_x)  {\psi_{L
                              \uparrow} })],
        	\end{equation}
        	where the terms in parentheses represent Kramer's pair at
                the two  edges  of the  system.\\
        The   authors    of   Ref
        \cite{Sarkar-2016,Altland-2011} have  mapped this  Hamiltonian--with Forward and umklapp interactions and in the proximity of s-wave superconductor ($\Delta$) and the
        magnetic field ($B$)--to  the XYZ
        spin-chain model (up to a constant) i.e, $H_{XYZ}= \sum_i H_i$,
        where
	\begin{equation} H_i = \sum_{\alpha} J_{\alpha} {S_i}^{\alpha} {S_{i+1}}^{\alpha}
	- [ \mu + B (-1)^i ] {S_i}^z .\end{equation} From bosonization
        procedure, fermionic field of 1D  quantum many body system can
        be      expressed      as,      $\psi_{R/L,\uparrow}(x)      =
        \frac{1}{2\pi\alpha}         n_{R,\uparrow}        e^{i\sqrt{4
            \pi}\phi_{R,\uparrow/\downarrow}(x)}$, where  $n_{L/R}$ is
        the  Klein  factor  to  preserve the  anticommutivity  of  the
        fermionic  field.   Here  we  introduce  two  bosonic  fields,
        $\theta(x)$ and  $\phi(x)$ which are  dual to each  other. The
        relations of  these two fields  are, $\phi(x) =  \phi_{R}(x) +
        \phi_{L}(x)$ and $\theta(x)  = \theta_{R}(x) + \theta_{L}(x)$.
         After
        the continuum field theory one can write the bosonised form of
        Hamiltonian as,
	\begin{equation}
	\begin{split}
	H = \int dx \frac{v}{2} \left[  \frac{1}{K} ( {({\partial_x \phi })}^2
          + K {({\partial_x \theta  })}^2 )\right] -\left( \frac{\mu}{\pi}\right)  \int dx 
        \partial_x  \phi +\left( \frac{B}{\pi}\right) \int dx   cos(\sqrt{4  \pi} \phi)\\  -
       \left( \frac{\Delta}{\pi}\right)  \int dx  cos(\sqrt{4  \pi }\theta)  +\left( \frac{g_u}{2
                 {\pi}^2 }\right)  \int dx  cos(4 \sqrt{\pi} \phi) ,
	\end{split}  \label{eqn:helical}
	\end{equation}
	 This  is   our  model  Hamiltonian   where,  $J_x=v_F+\Delta$,
	        $J_y=v_F-\Delta$ and $J_z=g_u$  are coupling constants. $\theta(x)$ and $\phi(x)$ are  the dual fields and $v = v_F
        + \frac{g_4}{2\pi}$, $v$ is the collective velocity and $v_F$ 
        is the Fermi velocity with $K = 1
        -\frac{g_2}{2\pi v_F};$ $K$ is the Luttinger liquid  parameter of
        the system.\\  Before we  begin to  discuss the  appearance of
        quantum BKT transition  in our system, we  discuss briefly why
        it is necessary  to study the quantum BKT  transition. Here we
        study two different situations for our model Hamiltonian.  (i)
        the proximity  induced superconducting  gap term is  absent ($
        \Delta  =0$) and  (ii) the  applied magnetic  field is  absent
        ($B=0 $).  For  both of these cases  only sine-Gordon coupling
        term is  present. Therefore,  there is no  competition between
        the  two mutually  non local  perturbation. Therefore  one can
        think that  there is no  need to study  the RG to  extract the
        quantum phases and phase boundaries.  But RG method is adopted
        for the  following reason. Each of  these Hamiltonians consist
        of two parts. The first part is the non-interacting term
        where  the  $\phi$  and  $\theta$ fields  show  the  quadratic
        fluctuations and the  other part of these  Hamiltonians is the
        sine-Gordon coupling  terms of either  of $\theta$ or  $\phi $
        fields. The  sine-Gordon coupling  term lock the  field either
        $\theta$   or  $\phi$   in   the  minima   of  the   potential
        well.  Therefore  the system  has  a  competition between  the
        quadratic part of the Hamiltonian and the sine-Gordon coupling
        term and this  competition will govern the  low energy physics
        of   these   Hamiltonians   in    different   limit   of   the
        system.\\\\
         \textbf{Quantum BKT equations}\\
	\noindent    We   consider    the   model    Hamiltonian   $H$
        (Eq.\ref{eqn:helical}) with $g_u=0$ since  it has no effect on
        the  topological state  and also  on  the Ising  state of  the
        system \cite{Sarkar-2016},
	\begin{equation}
		\begin{split}
		H  = \int dx \frac{v}{2}  \left[  \frac{1}{K} (  {({\partial_x
                      \phi  })}^2  +   K  {({\partial_x  \theta  })}^2
                  )\right]   -  \left(   \frac{\mu}{\sqrt{\pi}}\right)\int dx
                \partial_x  \phi  + \left( \frac{B}{\pi}\right) \int dx    cos(\sqrt{4  \pi}
                \phi)\\ - \left( \frac{\Delta}{\pi}\right)  \int dx cos(\sqrt{4 \pi \theta})
                \label{H}
		\end{split}
		\end{equation}
	The  Hamiltonian   under  the  situation,   proximity  induced
        superconducting gap term  is present, i.e, $  \Delta \ne0$ but
        the magnetic field is absent. We have,
	\begin{equation} 
	H_1  = \int dx \frac{v}{2}\left[  \frac{1}{K} (\partial_x  \phi)^2  +
          K(\partial_x    \theta)^2\right]     -   \left( \frac{\Delta}{\pi}\right)  \int dx
        \cos(\sqrt{4\pi}\theta(x)). \label{H1}
	\end{equation}
	The RG equation for this Hamiltonian can be derived as (we refer to appendix A for detailed derivation)
		\begin{equation}
			\dfrac{d{\Delta}}{dl}=  \left( 2-\frac{1}{K}\right)  {\Delta},
		        \;\;\;\;\;\;\; \frac{d K}{dl} = {\Delta}^2 . \label{RG1}
			\end{equation}  
		Under the situation that the applied magnetic field is present,
        i.e, $B\ne 0 $ but $\Delta$ is absent. The Hamiltonian is
	\begin{equation}
	H_2  = \int dx \frac{v}{2}\left[   \frac{1}{K}(\partial_x  \phi)^2  +
          K(\partial_x      \theta)^2\right]  + \left( \frac{B}{\pi}\right) \int dx 
        \cos(\sqrt{4\pi}\phi(x)). \label{H2}
	\end{equation}
	Following the same procedure as in appendix A, one can derive another
        set of quantum BKT equation,
	\begin{equation}
	\dfrac{dB}{dl}= (2-K)B, \;\;\;\;\; \frac{d K}{dl} = - B^2 K^2. \label{RG2}
	\end{equation}
	Thus,  we derive  two sets  of RG  equations.\\\\
\textbf{Length
          Scale Dependent Study of Quantum BKT transition: Evidence of
          Asymptotic Freedom like behavior}\\
	\noindent 
In this  section, we show explicitly how the physics of topology appears 
at different
length scales of the system to get further 
insight into the RG results and also obtain the signature of asymptotic 
freedom like 
behaviour.\\ 
        Asymptotic  freedom  is  a
        feature  of  QCD,  the  quantum field  theory  of  the  strong
        interaction between quarks and gluons
        \cite{PhysRevLett.30.1343,wu1991relativistic}. In QCD, the
        gauge theory of  quarks and  gluons are  asymptotically free,
        i.e.,  the coupling  vanishes  at very  short distance  (large
        momentum) and  grows at large distance  (small momentum). This
        allowed  us  to  understand  why  quarks  seemed  free  inside
        nucleons in  deep inelastic scattering and  are also confined
        at large distance.\\ 
        But the present  problem is not QCD. At the
        same  time we  are not  interpreting our  results in  terms of
        quark  and  gluon  physics,  rather in  terms  of  topological
        quantum  state of  matter. Therefore  we should  interpret our
        results  from the  length scale  dependent asymptotic  freedom
        like  behavior (not  the  strict asymptotic  freedom of  QCD).
        Length  scale  dependent  study  brings  out  the  concept  of
        asymptotic freedom  like behavior  in the  RG flow  sense. 
        \begin{figure}[]
        			\subfloat[]{\includegraphics[scale=0.45]{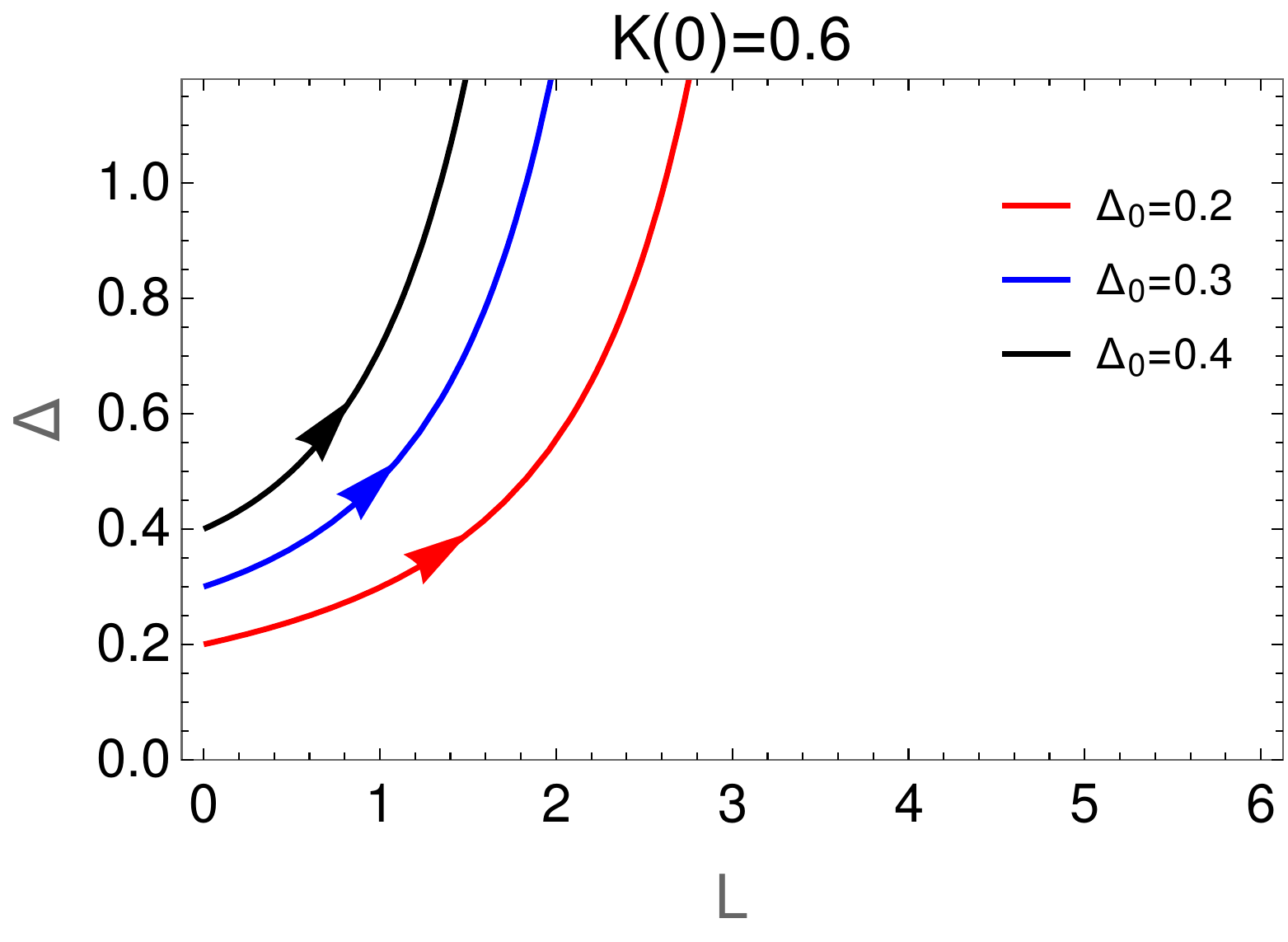}\label{figure3(a)}}
                                \;\;\;\subfloat[]{\includegraphics[scale=0.45]{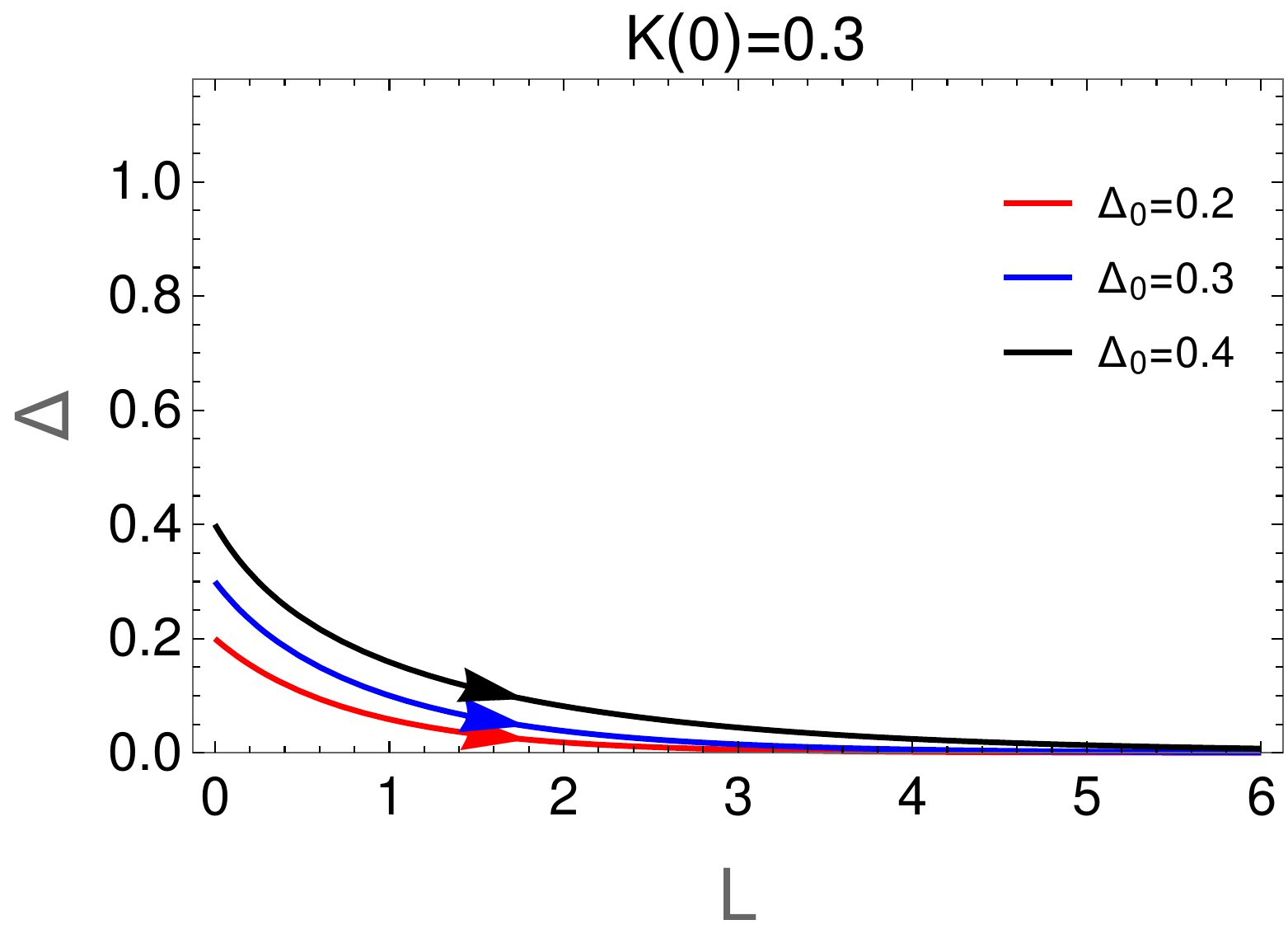}\label{figure3b}}\\
        			\caption{ This  figure consists of  two panels
                                  (a: $K=0.6$  and  b: $K=0.3$).  It   shows  the  variation  of
                                  $\Delta$  with $L$ for the  different initial
                                  values of $\Delta$.}\label{figure3}
        		\end{figure} 
        		\begin{figure}[] 
        					{\includegraphics[scale=0.5]{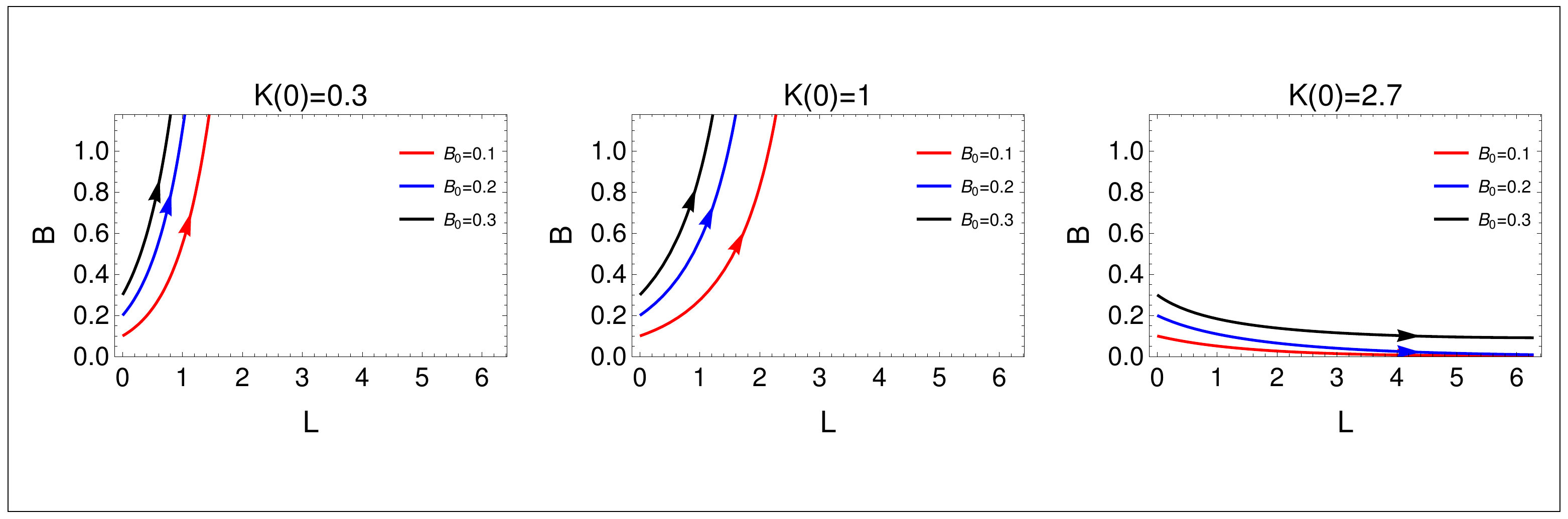}}
        							\caption{This figure shows the variation of $B$ with
        								$L$   for   different   initial   values   of
        								$B$.}\label{leng}
        						\end{figure}
        We
        would  like  to  explain  it explicitly  through  the  $\beta$
        function explanation. This asymptotic  freedom is a feature of
        all RG flows with a marginally relevant perturbation. This can
        be written as $\beta_{\lambda} =  C \lambda^2$, where $\lambda$ is the
        coupling constant and $C > 0$  is a constant 
(here
$\beta_{\lambda}$ is the $\beta$ function of the RG
theory from where one can predict the nature of the
RG flow lines of coupling constant $\lambda$). \\
In quantum field theory, flow lines are defined in energy scale but here we define RG
flow lines in length scale. With the model Hamiltonian we study here, we
study only the low energy properties. If we are interested in the high energy
behaviour of the system, how the coupling constant behave with the
larger momentum, i.e., for the small length scale. The physics of ultraviolet
behaviour occurs for that case only. Present study
involve the low energy physics for higher length scale. \\
\noindent 
In fig. \ref{figure3(a)}, we observe 
that RG flow lines flowing off to the strong coupling 
phase, i.e., the coupling $\Delta$ increases with the length scale. 
This coupling term induce the topological
superconducting phase in the system as we mentioned during
the introduction of the model Hamiltonian system. At the same time,
increasing $\Delta$ with the length scale is the signature of 
asymptotic freedom of the system, as we have discussed at the beginning of this section. Therefore the
topological superconducting phase appears in the asymptotic
degrees of freedom.\\    
In fig. \ref{figure3b} ($ K=0.3$, i.e, the system is in the
more repulsive regime),
we observe  that the coupling $\Delta$ decreases
rapidly with length scale and satisfies the condition for the
absence of the Majorana modes i.e., $L
\frac{\Delta}{v} <<1  $, ($L$ is the length, $v$ is
the collective velocity for the model Hamiltonian system) 
which shows the  system to be
in non-topological state  \cite{Sarkar-2016}. In this
phase $\Delta$ decreases rapidly with the length scale, i.e, there is
no asymptotic freedom here. For this situation the system shows the gapless helical
Luttinger liquid phase.\\
Therefore this study reveals that $K$ has a significant effect on  
the topological state of the system, as the values of $K$ become lower, i.e,
the system is in more strongly correlated phase which oppose the topological 
superconducting phase in the system.\\
Fig. 2 consists of three panels: the left, middle and right are
respectively for $K=0.3, 1$ and $2.7$. The
left and middle panels always show the asymptotic freedom like behaviour
for the coupling constant $B$. For these situations, $B$ induces the
Ising phase in the system, hence   
there is no
topological phase in the system. The system is
in the Ising phase for the asymptotic freedom regime otherwise it is
gapless helical Luttinger liquid phase.\\
Therefore it become clear from this study that $K$ has a significant
effect to change the asymptotic freedom like behaviour to the non-asymptotic
freedom like behaviour. The appearence of gapless helical Luttinger liquid phase for the
non-asymptotic freedom like behaviour regime is the same for the two sets of
RG equations (eq. \ref{RG1} and eq. \ref{RG2}). \\      
Thus it is clear from this study that for higher values of
$K (g_2  < 0,  K=1-\frac{g_2}{\pi v_F})$ the system is in the 
attractive regime. The system
transits from the asymptotic freedom like behavior to
gapless  helical Luttinger liquid  phase for the RG eq. \ref{RG2},
whereas the system is in non-asymptotic freedom like behaviour 
for strong repulsive region. Therefore it has become clear from this study that the concept of asymptotic freedom not only belongs to the high energy physics but alike behavior can also be observed in topological state of quantum matter system. \\\\
\textbf{Length
Scale Dependent Study of Quantum BKT Transition for
Finite $\mu$}\\
	The quantum BKT equations of $H_1$ (eq. \ref{H1}) for finite $\mu$ are,
	\begin{equation}
	\frac{d\Delta}{dl}  =   \left[  2  -  \frac{1}{K}\left(   1  +
          \frac{\mu}{v\pi}\right)   \right]   \Delta,   \;\;\;\;\;\;\;
        \frac{dK}{dl} = \Delta^2 .
	\end{equation}
	After doing the following  transformation, $-y_{||} = \left[ 2
          - \frac{1}{K}\left(  1  + \frac{\mu}{v\pi}\right)  \right]$,
        the BKT equations reduce to
	\begin{equation}
	\frac{d\Delta}{dl}   =    -y_{||}   \Delta,   \;\;\;\;\;\;\;\;
        \frac{dy_{||}}{dl}        =       -\frac{\Delta^2}{(1        -
          \frac{\mu}{v\pi})}. \label{RGdeltamu}
	\end{equation}
There will be no corrections in the RG equation for the
Hamiltonian $H_2$ (eq. \ref{H2}). It is because the sine-Gordon coupling is also for $\phi$ field. For this situation one
can absorb the chemical potential term in the sine-Gordon coupling [3].
	\begin{figure}[H]
		\subfloat[]{\includegraphics[scale=0.5]{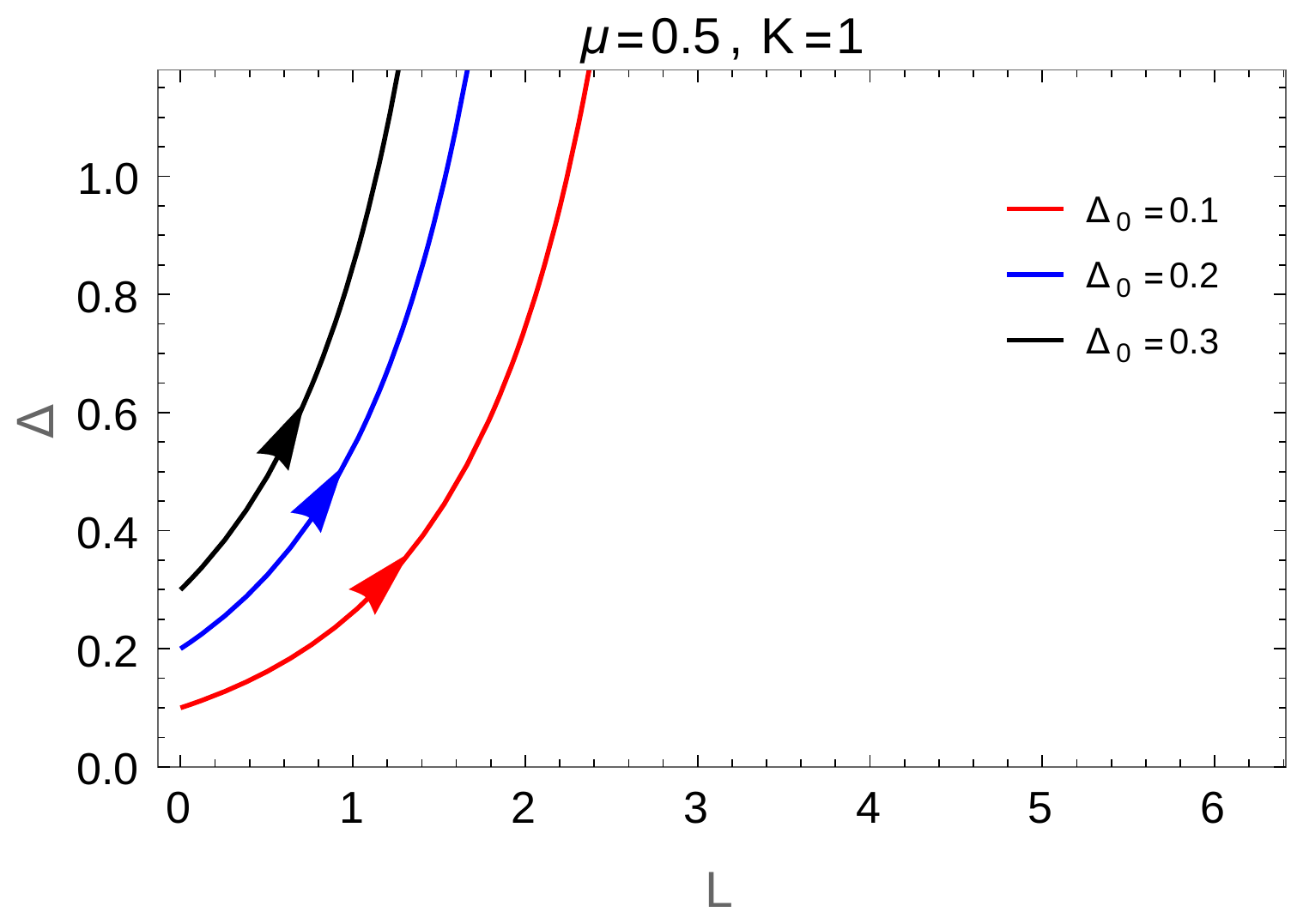}\label{figure5b}}
                \;\;\;\subfloat[]{\includegraphics[scale=0.5]{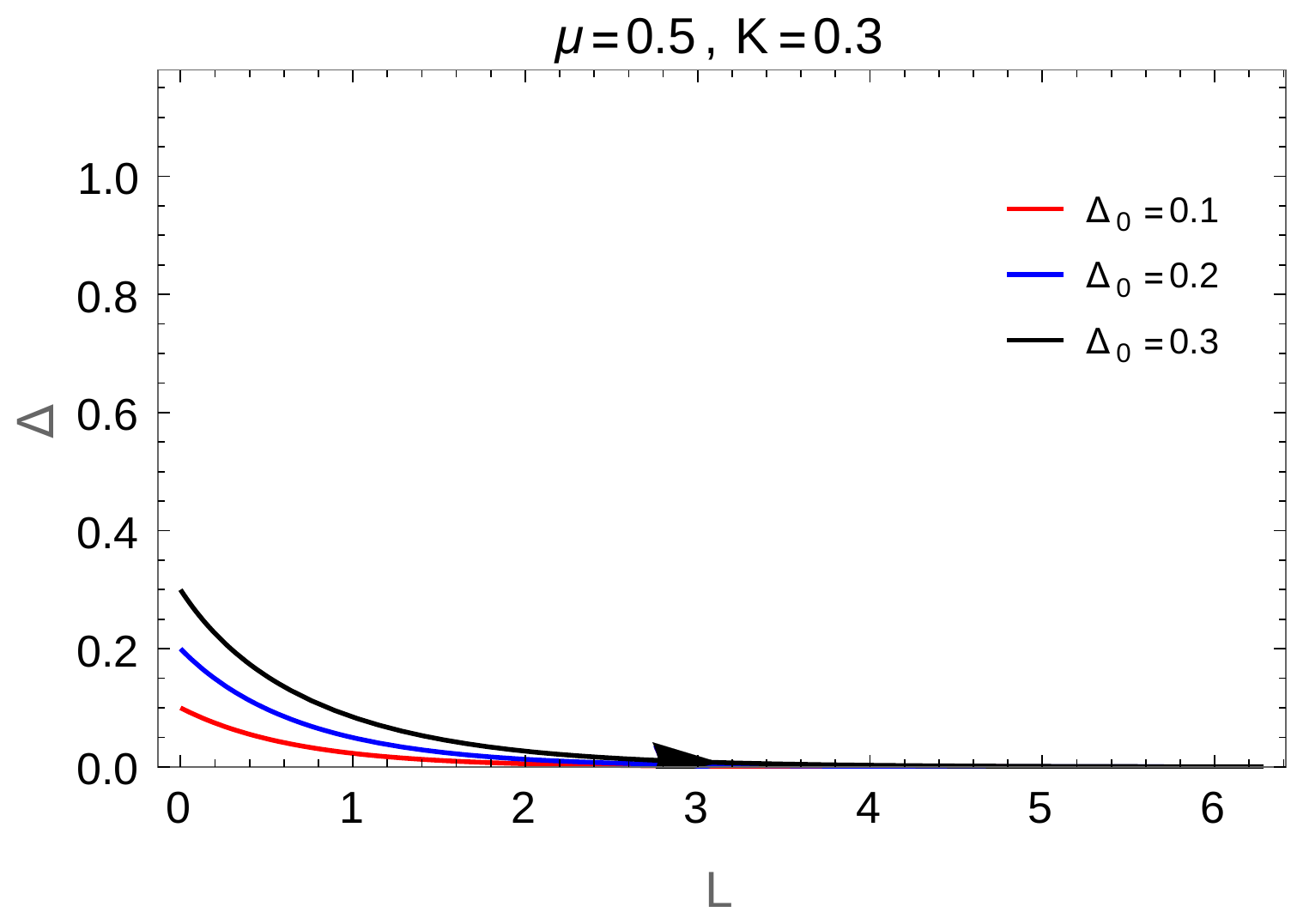}\label{figure5a}}
		 \caption{This figure  consists of  two panels  (a: $\mu =0.5$ and b: $\mu =1 $
                   ). It shows  the variation of $\Delta$  with L for
                   the different initial values of
                   $\Delta$.}\label{figure5}
	\end{figure} 
Fig. 3 shows the results for finite $\mu$ (eq. $51$). 
This figure consists of two panels (a and b), left and right, for 
$K=1 $  and $0.3$ respectively, for the fixed value $\mu =0.5$. The different 
curves in each
figure are for the different initial values of $\Delta $. We observe
for fig.\ref{figure5b} ($\mu =0.5 $ \& $K=1$), RG flow lines flowing off to the strong coupling 
phase with length scale and finally the system is in the 
topological superconducting phase. In fig.\ref{figure5a} we observe the RG flow lines decreasing with the length scale which indicates the system is in the  gapless helical Luttinger liquid phase. Thus we conclude that the presence of $\mu$ does not have any effect on the asymptotic freedom like behavior of the system.\\\\
{\bf Summary of the new and important results of the present study 
:}\\
We have obtained three quantum phases in this length scale 
dependent RG flow diagram study.
One is  topological, i.e., topological superconducting phase, another one is the 
Ising phase which is non-topological and finally we have obtained gapless 
helical Luttinger liquid phase which is also non-topological in character. Luttinger liquid parameter $K$, has played most    important role to achieve these quantum phases. We have observed that the asymptotic freedom like behaviour can be present both in topological (superconducting phase) and non-topological (Ising phase) phases. 
We have studied the effect of finite $\mu$ and observed the behavior of the asymptotic nature is intact.
To the best of our knowledge this new and
important results are absent in the literature of topological state of matter. This is the first study where we show the presence of asymptotic freedom like behavior in the area of condensed matter systems with topological background even though it was only observed in the area high energy physics. \\\\
{\bf Plausible experimental verification : }\\
One can find this length scale dependent quantum phases, 
which we presented in the previous section from the following 
consideration. The first step is to quantum simulate the topological
quantum state in semiconductor nanowire for different 
length scale, i.e, hybrid superconductor-semiconductor nanowire \cite{exp-MF}. One has to  very
fine control on the proximity induced superconductivity gap ($\Delta$) and the
applied magnetic field ($B$), these are the two key players for the different
quantum phases either topological or non-topological in character. The other points
to be noted that the $K$ and $\mu$  must have to be vary with a
high precision either by using the gate voltage or by the doping manipulation.
The nanowires of different length scale with variation of these four factors
can achieve the all phases and transition among themselves.\\\\
\noindent \textbf{Conclusion}\\
We  have done the  detailed length scale dependent study of quantum  BKT transition
for  interacting helical  liquid  system at  the  edge of  topological
insulator. We have  found the presence of asymptotic  freedom like behavior for
the  topological state  of  the system  for  finite proximity  induced
superconducting gap  and also for the Ising  phase of  the system  in the
presence  of magnetic  field. There is no evidence of superconductor-insulator transition in this asymptotic freedom study. We have presented importance of the 
Luttinger liquid parameter in different phases of the study. We have observed
the presence of helical Luttinger liquid phase for the two different
BKT transitions as a common quantum phase. We have also presented the
results for finite chemical potential.
This work provides a new perspective on the study of the topological state of 
quantum matter.\\\\
\textbf{Acknowledgment}\\
\noindent R.K.R and R.S would  like to acknowledge 
Mr. N. Prakash and
Prof. R. Srikanth for reading the manuscript critically. R.K.R and R.S also acknowledge 
RRI library  for the books and  journals and ICTS for
Lectures/seminars/workshops/conferences/discussion meetings of
different aspects of physics. S.S would  like to acknowledge
DST (EMR/2017/000898)
for the  support. \\

	\appendix

\textbf{A. Detailed derivation of RG equation for $H_1$ Hamiltonian}\\
In  the  Bosonized model  Hamiltonian  $H_1$ is written as
\begin{equation} 
	H_1  = \int dx \frac{v}{2}\left[  \frac{1}{K} (\partial_x  \phi)^2  +
          K(\partial_x    \theta)^2\right]     -   \left( \frac{\Delta}{\pi}\right)  \int dx
        \cos(\sqrt{4\pi}\theta(x)). 
	\end{equation}
	 We rescale  the
	        fields as, $\phi  \rightarrow \phi^{\prime}=\phi/\sqrt{K}$ and
	        $\theta \rightarrow  \theta^{\prime}=\sqrt{K}\theta$. 
	        The Lagrangian for quadratic and interaction terms can be written as
		\begin{equation}
		\mathcal{L}_0=
	        -\frac{1}{4}[v^{-1}(\partial_{\tau}\theta^{\prime})^2  +
	          v(\partial_{x}\theta^{\prime})^2 ] ; \;\;\;\;\;\;
		\mathcal{L}_{\Delta}     =     \left(\frac{\Delta}{\pi}\right)
	        \cos(\sqrt{4\pi}\theta(x)) ,
		\end{equation} 
		where $\tau = it$ is the imaginary time.
		The  Euclidean  action  can  be written  as,  $S_E=  -\int  dr
	        \mathcal{L}      =     -\int      dr     (\mathcal{L}_0      +
	        \mathcal{L}_{\Delta})$, where  $r=(\tau,x)$. The
	        partition function in terms of Euclidean action as
		\begin{equation}
		\mathcal{Z}= \int \mathcal{D}\theta^{\prime} e^{\left[ \int dr
	            \left(-\frac{1}{4}v^{-1}(\partial_{r}\theta^{\prime})^2
	            -\frac{1}{4}   v(\partial_{x}\theta^{\prime})^2\right)   -
	            \int dr \mathcal{L}_{\Delta}(\theta^{\prime}) \right]}.
		\end{equation}
		\begin{equation}
		\mathcal{Z}=       \int      \mathcal{D}\theta       e^{\left[
	            -\int_{-\Lambda}^{\Lambda}    \frac{d\omega}{2\pi}|\omega|
	            \frac{K|\theta(\omega)|^2}{2}       -        \int       dr
	            \mathcal{L}_{\Delta}(\theta)\right]}.
		\end{equation}
		We now separate the
	        slow  and  fast  fields  and  integrate  out  the  fast  field
	        components.        The         field        $\theta$        is
	        $\theta(r)=\theta_s(r)+\theta_f(r)$ where,
		\begin{equation}
		\theta_s(r)=\int_{-\Lambda/b}^{\Lambda/b} \frac{d\omega}{2\pi}
	        e^{-i\omega\tau}\theta(\omega)                \;\;\;\;\&\;\;\;
	        \theta_f(r)=\int_{\Lambda/b<|\omega_n|<\Lambda}
	        \frac{d\omega}{2\pi} e^{-i\omega\tau}\theta(\omega),
		\end{equation}
		here $r=(x,\tau)$.  Now the partition function  can be written
	        as,
		\begin{equation}
		\begin{aligned}
		\mathcal{Z}&=\int  \mathcal{D}\theta_s \mathcal{D}\theta_f  e^
	                {[-S_s(\theta_s)-S_f(\theta_f)-
	                    S_{\Delta}(\theta_s,\theta_f)]},\\     &=     \int
	                \mathcal{D}\theta_s e^ {[-S_s(\theta_s)]} \left\langle
	                e^{[- S_{\Delta}(\theta_s,\theta_f)]}\right\rangle_f,
		\end{aligned}
		\end{equation}
		The effective action can be written as cumulant expansion up to second order as
		\begin{equation}
		S_{eff}(\theta_s)=     S_s(\theta_s)     -     \ln\left\langle
	        e^{-S_{\Delta}(\theta)} \right\rangle_f.
		\end{equation}
		\begin{equation}
		S_{eff}(\theta_s)=                  S_s(\theta_s)+\left\langle
	        S_{\Delta}(\theta_s,\theta_f)\right\rangle                   -
	        \frac{1}{2}(\left\langle
	        S_{\Delta}^2(\theta_s,\theta_f)\right\rangle-\left\langle
	        S_{\Delta}(\theta_s,\theta_f)\right\rangle^2).
		\end{equation}
		Now we  calculate the first order  approximation $\left\langle
	        S_{\Delta}(\theta_s,\theta_f)\right\rangle$,
		\begin{equation}
		\begin{aligned}
		\left\langle    S_\Delta(\theta_s,\theta_f)\right\rangle    &=
	        \frac{\Delta}{\pi}\int\mathcal{D}\theta_f
	        e^{-S_f(\theta_f)}\int             dr             \left\langle
	        \cos(\sqrt{4\pi}\theta(r))\right\rangle,\\                  &=
	        \frac{\Delta}{2\pi}\int                         dr\left\lbrace
	        e^{i\sqrt{4\pi}\theta_s(r)}  \int\mathcal{D}\theta_f e^{\left(
	          \int_f\frac{d\omega}{2\pi}[i\sqrt{4\pi}e^{i\omega
	              r}\theta_f-|\omega|      \frac{K|\theta_f|^2}{2}]\right)
	        }+H.c   \right\rbrace   ,\\   &=   \frac{\Delta}{\pi}\int   dr
	        \cos[\sqrt{4\pi}\theta_s(r)]  e^{\left(   -\frac{1}{K}  \int_f
	          \frac{d\omega}{|\omega|} \right) }.\\
		\end{aligned}
		\end{equation}
		We                        write,                       $\int_f
	        \frac{d\omega}{|\omega|}=\int_{\Lambda/b}^{\Lambda}
	        \frac{d\omega}{|\omega|}=
	        \ln\Lambda-\ln(\Lambda/b)=\ln[\frac{\Lambda}{\Lambda/b}]=\ln
	        b$.
		\begin{equation}
		\begin{aligned}
		\left\langle    S_\Delta(\theta_s,\theta_f)\right\rangle    &=
	        \frac{\Delta}{\pi}\int     dr     \cos[\sqrt{4\pi}\theta_s(r)]
	        e^{\left( -\frac{1}{K} \ln b  \right) },\\ &= b^{-\frac{1}{K}}
	        S_{\Delta}(\theta_s).
		\end{aligned}
		\end{equation}
		Thus the effective action upto first order cumulant expansion
	        can be written as,
		$$   S_{eff}(\theta_s)=   S_s(\theta_s)   +   b^{-\frac{1}{K}}
	        S_\Delta(\theta_s),$$
		\begin{equation}
		S_{eff}(\theta_s)=               \int_{-\Lambda/b}^{\Lambda/b}
	        \frac{d\omega}{2\pi}|\omega| \frac{K|\theta_s(\omega)|^2}{2} +
	        b^{-\frac{1}{K}}  \int   dr  \left(  \frac{\Delta}{\pi}\right)
	        \cos[\sqrt{4\pi}\theta_s(r)].
		\end{equation}
		Now we rescale the parameters cut-off momentum to the original
	        momentum by considering, $\bar{\Lambda}=\frac{\Lambda}{b}$ , $
	        \bar{\omega}=\omega b$ and  $\bar{r}= \frac{r}{b}$. The fields
	        will    be     rescaled    as,    $\bar{\theta}(\bar{\omega})=
	        \frac{\theta(\omega)}{b}$         and        we         choose
	        $\bar{\theta}(\bar{r})=\theta_s(r)$.  Also since the system is  (1+1)  dimensional we  have
	                $d^2r= b^2d^2\bar{r}$. Thus   the    rescaled
	        effective action is given by,
		\begin{equation}
		\begin{aligned}
		S_{eff}(\theta_s)&=                  \int_{-\Lambda}^{\Lambda}
	        \frac{d\bar{\omega}}{2\pi           b}\frac{\bar{|\omega|}}{b}
	        \frac{b^2K|\bar{\theta}(\bar{\omega})|^2}{2}                 +
	        b^{-\frac{1}{K}}      \int      b^2     d^2\bar{r}      \left(
	        \frac{\Delta}{\pi}\right)
	        \cos[\sqrt{4\pi}\bar{\theta}(\bar{r})],\\                   &=
	        \int_{-\Lambda}^{\Lambda}      \frac{d\bar{\omega}}{2\pi     }
	        \bar{|\omega|}   \frac{K|\bar{\theta}(\bar{\omega})|^2}{2}   +
	        b^{2-\frac{1}{K}}        \int         d^2\bar{r}        \left(
	        \frac{\Delta}{\pi}\right)
	        \cos[\sqrt{4\pi}\bar{\theta}(\bar{r})].
		\end{aligned}
		\end{equation}
		Comparing the coupling constants  of rescaled effective action
	        with  the   not renormalized  action   one  can   observe  that,
	        $\Delta\rightarrow  \Delta b^{2-\frac{1}{K}}$.  Thus we  write
	        the   RG   flow   equation    as,   $   \bar{\Delta}=   \Delta
	        b^{2-\frac{1}{K}}.$  We   write   this  equation   in   the
	        differential form defining    the
	                differential of  a parameter  as $d\Delta=\bar{\Delta}-\Delta$ and setting $b=e^{dl}$,
		\begin{equation}
		\dfrac{d\Delta}{dl}= \left( 2-\frac{1}{K}\right) \Delta.
		\end{equation}      
		Now we solve for the second order cumulant expansion,
		\begin{dmath}
	        -\frac{1}{2}(\left\langle
			 S_{\Delta}^2\right\rangle-\left\langle
			 S_{\Delta}\right\rangle^2) =-\frac{{\Delta}^2}{2\pi^2}
			- \int    dr     dr^{\prime}    \left[    \left\langle
			 \cos[\sqrt{4\pi}\theta(r)]\right\rangle \left\langle
			 \cos[\sqrt{4\pi}\theta(r^{\prime})]    \right\rangle
			- \left\langle              \cos[\sqrt{4\pi}\theta(r)]
			 \right\rangle                           \left\langle
			 \cos[\sqrt{4\pi}\theta(r^{\prime})]    \right\rangle
			 \right].\end{dmath}
		First we calculate $\left\langle S_{\Delta}^2\right\rangle $ term. 
		\begin{dmath*}
			\left\langle
	                S_{\Delta}^2\right\rangle=\frac{{\Delta}^2}{\pi^2}\int
	                dr dr^{\prime} \left\langle \cos[\sqrt{4\pi}\theta(r)]
	                \cos[\sqrt{4\pi}\theta(r^{\prime})] \right\rangle ,
		\end{dmath*}
		\begin{dmath}
			\left\langle        S_{\Delta}^2\right\rangle        =
	                \frac{{\Delta}^2}{2\pi^2}\int  dr  dr^{\prime}  \left[
	                  \cos\sqrt{4\pi}[\theta_s(r)+\theta_s(r^{\prime})]
	                  \left(  e^{-2\pi\left[   \left\langle  \theta_f^2(r)
	                      \right\rangle+                      \left\langle
	                      \theta_f^2(r^{\prime})      \right\rangle      +
	                      2\left\langle    \theta_f(r)\theta_f(r^{\prime})
	                      \right\rangle\right]         }\right)\\        +
	                  \cos\sqrt{4\pi}[\theta_s(r)-\theta_s(r^{\prime})]
	                  \left(  e^{-2\pi  \left[ \left\langle  \theta_f^2(r)
	                      \right\rangle+                      \left\langle
	                      \theta_f^2(r^{\prime})      \right\rangle      -
	                      2\left\langle    \theta_f(r)\theta_f(r^{\prime})
	                      \right\rangle\right] }\right)+H.c \right].
		\end{dmath}
		Now we calculate $\left\langle S_{\Delta}\right\rangle^2$,
		
		\begin{dmath*}
			\left\langle
	                S_{\Delta}\right\rangle^2=\frac{{\Delta}^2}{\pi^2}\int
	                dr dr^{\prime} \left\langle \cos[\sqrt{4\pi}\theta(r)]
	                \right\rangle                             \left\langle
	                \cos[\sqrt{4\pi}\theta(r^{\prime})] \right\rangle ,
		\end{dmath*}
		
		
		\begin{dmath}
			\left\langle        S_{\Delta}\right\rangle^2        =
	                \frac{{\Delta}^2}{2\pi^2}\int  dr  dr^{\prime}  \left[
	                  \cos\sqrt{4\pi}[\theta_s(r)+\theta_s(r^{\prime})]
	                   e^{-2\pi\left\langle \theta^2_f(r)\right\rangle }
	                  e^{-2\pi\left\langle
	                    \theta^2_f(r^{\prime})\right\rangle       }      \\+
	                  \cos\sqrt{4\pi}[\theta_s(r)-\theta_s(r^{\prime})]
	                  e^{-2\pi\left\langle   \theta^2_f(r)\right\rangle  }
	                  e^{-2\pi\left\langle
	                    \theta^2_f(r^{\prime})\right\rangle }+ H.c \right]
	                .
		\end{dmath}
		Thus            the             term            $(\left\langle
	        S_{\Delta}^2\right\rangle-\left\langle
	        S_{\Delta}\right\rangle^2)$ is,
		\begin{dmath}
			(\left\langle   S_{\Delta}^2\right\rangle-\left\langle
	          S_{\Delta}\right\rangle^2)= \frac{{\Delta}^2}{2\pi^2}\int dr
	          dr^{\prime}                                           \left[
	            \cos\sqrt{4\pi}[\theta_s(r)+\theta_s(r^{\prime})]   \left(
	            e^{-2\pi\left[  \left\langle \theta_f^2(r)  \right\rangle+
	                \left\langle  \theta_f^2(r^{\prime})  \right\rangle  +
	                2\left\langle          \theta_f(r)\theta_f(r^{\prime})
	                \right\rangle\right]             }\right)            +
	            \cos\sqrt{4\pi}[\theta_s(r)-\theta_s(r^{\prime})]   \left(
	            e^{-2\pi \left[  \left\langle \theta_f^2(r) \right\rangle+
	                \left\langle  \theta_f^2(r^{\prime})  \right\rangle  -
	                2\left\langle          \theta_f(r)\theta_f(r^{\prime})
	                \right\rangle\right]    }\right)+   H.c    \right]   -
	          \frac{{\Delta}^2}{2\pi^2}\int    dr    dr^{\prime}    \left[
	            \cos\sqrt{4\pi}[\theta_s(r)+\theta_s(r^{\prime})]
	            e^{-2\pi\left\langle      \theta^2_f(r)\right\rangle     }
	            e^{-2\pi\left\langle \theta^2_f(r^{\prime})\right\rangle }
	            +        \cos\sqrt{4\pi}[\theta_s(r)-\theta_s(r^{\prime})]
	            e^{-2\pi\left\langle      \theta^2_f(r)\right\rangle     }
	            e^{-2\pi\left\langle   \theta^2_f(r^{\prime})\right\rangle
	            }+ H.c \right] ,
		\end{dmath}
		\begin{dmath}
		-\frac{1}{2}(\left\langle
	        S_{\Delta}^2\right\rangle-\left\langle
	        S_{\Delta}\right\rangle^2)  = -\frac{{\Delta}^2}{4\pi^2}  \int
	        dr              dr^{\prime}             \\              \left[
	          \cos\sqrt{4\pi}[\theta_s(r)+\theta_s(r^{\prime})]
	          e^{-2\pi[\left\langle    \theta_f^2(r)    \right\rangle    +
	              \left\langle    \theta_f^2(r^{\prime})   \right\rangle]}
	          \left(  e^{-4\pi\left\langle \theta_f(r)\theta_f(r^{\prime})
	            \right\rangle         }        -1         \right)        +
	          \cos\sqrt{4\pi}[\theta_s(r)-\theta_s(r^{\prime})]
	          e^{-2\pi[\left\langle    \theta_f^2(r)    \right\rangle    +
	              \left\langle    \theta_f^2(r^{\prime})   \right\rangle]}
	          \left(  e^{4\pi\left\langle  \theta_f(r)\theta_f(r^{\prime})
	            \right\rangle } -1 \right) \right] . \label{seconddelta}
		\end{dmath}
		The         correlation         function         $\left\langle
	        \theta_f(r)\theta_f(r^{\prime})  \right\rangle$ for $r^{\prime}\rightarrow r$ can be written as \cite{A.Storm-paper},
		\begin{equation}
		\left\langle   \theta_f(r)\theta_f(r^{\prime})   \right\rangle
	        \approx     \left\langle      \theta_f^2(r)\right\rangle     =
	        \frac{1}{2\pi                     K}\int_{\Lambda/b}^{\Lambda}
	        \frac{d\omega}{\omega}= \frac{1}{2\pi K} \ln b .
		\end{equation}
		We  introduce  the  relative coordinate  $s=r-r^{\prime}$  and
	        center of mass coordinate  $T=(r+r^{\prime})/2$. For small $s$
	        cosine can be approximated to,
		\begin{equation*}
		\cos\sqrt{4\pi}[\theta_s(r)+\theta_s(r^{\prime})]=\cos
	                 [4\sqrt{\pi}(\theta_s(T))];                  \;\;\;\;
	                 \cos\sqrt{4\pi}[\theta_s(r)-\theta_s(r^{\prime})]=  1
	                 - 2\pi (s\partial_T\theta_s(T))^2.
		\end{equation*} 
		Thus eq.\ref{seconddelta} can be written as,
		\begin{equation}
		\begin{aligned}
		-\frac{1}{2}(\left\langle
	        S_{\Delta}^2\right\rangle-\left\langle
	        S_{\Delta}\right\rangle^2) =-\frac{\Delta^2}{4\pi^2}   \left(   1-    \left(\frac{1}{b}
	        \right)^{\frac{2}{K}} \right) \int_{0}^{b/\Lambda}  ds \int dT
	        ( 1 - 2\pi (s\partial_T\theta_s(T))^2).\\
		\end{aligned}
		\end{equation}
		Here  the  first  term  turns  out  to  be  field  independent
	        term. Thus we consider only second term,
		\begin{equation}
		\begin{aligned}
		-\frac{1}{2}(\left\langle S_{\Delta}^2\right\rangle-\left\langle S_{\Delta}\right\rangle^2)&
	=\frac{{\Delta}^2}{4\pi^2} \left( 1- \left(\frac{1}{b} \right)^{\frac{2}{K}} \right) 
	        \int_{0}^{b/\Lambda} ds 
	\int dT (2\pi (s\partial_T\theta_s(T))^2),\\
		&=\frac{{\Delta}^2}{2\pi} \left( 1- \left(\frac{1}{b} \right)^{\frac{2}{K}} \right) \int_{0}^{b/\Lambda} 
	s^2 ds \int dT  (\partial_T\theta_s(T))^2,\\
		&=\frac{{\Delta}^2}{6\pi\Lambda^3} \left( \left(\frac{1}{b} \right)^{-3}- 
	\left(\frac{1}{b} \right)^{\frac{2}{K}-3} \right) \int_{-\Lambda/b}^{\Lambda/b} 
	\frac{d\omega}{2\pi}|\omega| \frac{K|\theta_s(\omega)|^2}{2}.
		\end{aligned}
		\end{equation}
		After rescaling  the parameters  and fields the  equation will
	        have the form,
		\begin{equation}
		-\frac{1}{2}(\left\langle
	        S_{\Delta}^2\right\rangle-\left\langle
	        S_{\Delta}\right\rangle^2)=\frac{{\Delta}^2}{6\pi\Lambda^3}
	        \left(   \left(\frac{1}{b}   \right)^{-3}-   \left(\frac{1}{b}
	        \right)^{\frac{2}{K}-3}    \right)   \int_{-\Lambda}^{\Lambda}
	        \frac{d\bar{\omega}}{2\pi           }           \bar{|\omega|}
	        \frac{K|\bar{\theta}(\bar{\omega})|^2}{2}.
		\end{equation}
		Now the effective action can be written as, 
		\begin{equation}
		\begin{split}
			S_{eff}=                     \int_{-\Lambda}^{\Lambda}
	                \frac{d\bar{\omega}}{2\pi       }       \bar{|\omega|}
	                \frac{K|\bar{\theta}(\bar{\omega})|^2}{2}            +
	                b^{2-\frac{1}{K}}      \int     d^2\bar{r}      \left(
	                \frac{{\Delta}}{\pi}\right)
	                \cos[\sqrt{4\pi}\bar{\theta}(\bar{r})]\\ +\frac{{\Delta}^2}{6\pi\Lambda^3}
	                \left(         \left(\frac{1}{b}         \right)^{-3}-
	                \left(\frac{1}{b}    \right)^{\frac{2}{K}-3}   \right)
	                \int_{-\Lambda}^{\Lambda}  \frac{d\bar{\omega}}{2\pi }
	                \bar{|\omega|}
	                \frac{K|\bar{\theta}(\bar{\omega})|^2}{2},
		\end{split}
		\end{equation}
		\begin{equation}
		S_{eff}  =  \left[  1+\frac{{\Delta}^2}{6\pi\Lambda^3}  \left(
	          \left(\frac{1}{b}       \right)^{-3}-      \left(\frac{1}{b}
	          \right)^{\frac{2}{K}-3}   \right)   \right]   S_s(\theta_s)+
	        b^{2-\frac{1}{K}}S_{\Delta}(\theta_s).
		\end{equation}      
		Comparing  the rescaled  effective  action  with the  original
	        action we obtain RG flow equation,
		\begin{equation}
		{\bar{K}}=  K\left[ 1+\frac{{\Delta}^2}{6\pi\Lambda^3}  \left(
	          \left(\frac{1}{b}       \right)^{-3}-      \left(\frac{1}{b}
	          \right)^{\frac{2}{K}-3} \right) \right] .
		\end{equation}
		 Defining the  differential of  a parameter  as $dK=\bar{K}-K$
	         and  by setting  $b=e^{dl}$ we  get RG  flow equation  in the
	         differential form
		 $$                                                        dK=
	         -\frac{K\Delta^2}{6\pi\Lambda^3}(e^{3dl}-e^{(3-\frac{2}{K})dl}),$$
		 \begin{equation}
		 \dfrac{dK}{dl}=\left(\frac{{\Delta}^2}{3\pi\Lambda^3}\right).
		 \end{equation}
		We rescale ${\Delta}  \rightarrow
	        {\Delta}\sqrt{\frac{1}{3\pi\Lambda^3}}$. Thus RG flow
	        equations of $H_1$ are given by
		\begin{equation}
		\dfrac{d{\Delta}}{dl}=  \left( 2-\frac{1}{K}\right)  {\Delta},
	        \;\;\;\;\;\;\; \frac{d K}{dl} = {\Delta}^2 .
		\end{equation}  
		Similarly one can derive quantum BKT equation for Hamiltonian $H_2$ as
			\begin{equation}
			\dfrac{dB}{dl}= (2-K)B, \;\;\;\;\; \frac{d K}{dl} = - B^2 K^2. 
			\end{equation}
	\end{document}